Chapter 6

# Metacognition and self-regulated learning in manipulative robotic problem-solving task

Margarida Romero[1] and George Kalmpourtzis[1]

1 Laboratoire d'Innovation et Numérique pour l'Education, Université Côte d'Azur, France, https://orcid.org/0000-0003-3356-8121

Margarida.Romero@unice.fr, gkalmpourtzis@playcompass.com, https://orcid.org/0000-0003-0761-6441

## Abstract

Metacognition is an important aspect in creative problem solving (CPS) and through this chapter we analyse the meta-reasoning aspects applied in the different processes of monitoring the progress of learners' reasoning and CPS activities. Meta-reasoning monitors the way that problem-solving processes advance and regulate time and efforts towards a solution. In the context of an ill-defined problem, exploration is required to develop a better-defined problem space and advance towards the solution space. The way learners engage in exploration and exploitations is regulated by the meta-reasoning within the CPS activity. The objective of this chapter is to examine and identify the CPS process with educational robots through a metacognitive and interactionist approach. This chapter presents a case study, where, to solve a problem, a participant had to explore a set of robot cubes to develop the technological knowledge associated with each single component of the system, but also conceptualize a system-level behaviour of the cubes when they are assembled. The chapter presents the emergence of knowledge through the metacognitive regulation of the process of exploration and exploitation of prior knowledge and emergent knowledge until finding a solution.

## Introduction

Metacognition has been defined as the knowledge and regulation of one's own cognitive processes (Campione and Brown, 1987), an "intertwining of awareness (i.e., self-monitoring) and use of that awareness (i.e., self-regulation) to advance a process (e.g., writing, studying, learning transfer, driving, cooking)" according to Scharff and colleagues (Scharff et al., 2017). The metacognitive awareness and self-regulation are important processes to advance a problem-solving task (Berardi-Coletta et al., 1995; Davidson et al., 1994), especially in ill-defined or complex problem-solving tasks (Kim et al., 2013; Scharff et al., 2017). In this chapter we analyse creative problem-solving with an interactive manipulable under the lens of metacognition, in order to advance the understanding of creative problem-solving (CPS) tasks engaging manipulable "visuo-spatial constructive play objects" (VCPOs) (Ness and Farenga, 2016) including “blocks (for example, standard wood blocks, plastic blocks, and foam blocks), bricks (such as LEGO bricks and MegaBloks), and planks (1×3×15 cm wooden rectangular cuboids)” [p. 202]. In the problem-solving tasks with manipulable VCPOs as educational robotics there is a lack of studies addressing the specific metacognitive aspects which are engaged during the CPS with VCPOs modular educational robots. With this objective, we also analyse the emergence of knowledge through the metacognitive regulation of the process of exploration and exploitation of prior knowledge and emergent knowledge until finding a solution.

## Creative problem solving

Problem-solving engages learners in cognitive and metacognitive processes towards the accomplishment of specific goals. Metacognition is especially important in ill-structured problem-solving tasks (An and Cao, 2014), in which the learners do not dispose of all necessary knowledge and procedures to solve the presented tasks. While some problems can be solved through planning and executing a set of actions from an initial state towards the goal state (Newell and Simon, 1972), other types of problems require the exploration of the problem space and its different components to refine the initial situation and engage in a process of its construction before approaching a solution space. In creative problem-solving tasks, the solution space is composed by several potential goal states. Problem solving processes, during which the problem space and the solution space may be formed through different stages, can be considered as creative problem-solving (CPS). CPS is opposed to analytical problem solving, in which the application of an algorithm or a set of rules can lead to a solution. Furthermore, CPS “tends to be characterized by more divergent, associational or discontinuous solution processes” (Jarosz et al., 2012, p. 487). During the CPS process, metacognitive regulation is important to help adapt learners’ strategies in relation to the progression in problem-solving and the consideration of divergent and convergent problem-solving.

## An interactionist and metacognitive approach to problem solving

From an interactionist approach, creative behaviour is the result of the interaction between the person and the context (Woodman et al., 1993). As a result, in CPS, we should consider participants’ prior knowledge, the emergent knowledge developed during the task performance but also the state of the robotic components across the task. The emergence of knowledge through the exploration process leads to the evolution of participants’ knowledge but also to the actualization of the technological system they interact with: when using unfamiliar technologies to achieve a certain task, the participant confronts what Norman (Norman, 1986) designates as the *gulf of execution*, the distance between the user's goals and the means of achieving them through the system. The gulf of execution should be crossed through the exploration of the means or tools available to him to achieve a certain goal or solving a certain problem-situation (execution bridge) and evaluating the effect of these actions (evaluation bridge).

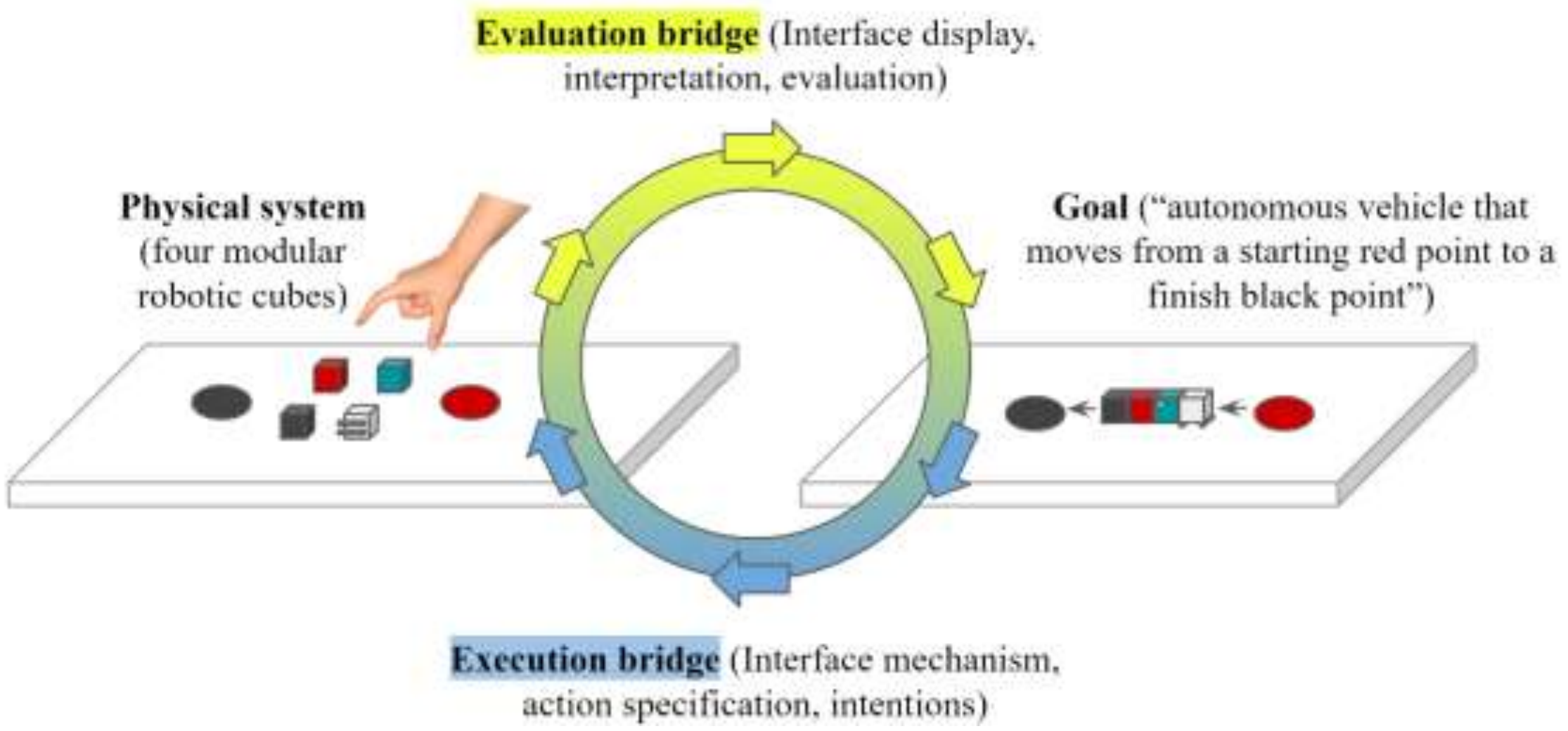

Figure 1 – Evaluation and execution bridges in the CreaCube task

During the process of overcoming the *gulf of execution* and creating a bridge between users' goals and the means towards achieving them, through a system, their mental and physical activity with the system creates a series of loops of interaction to actualize the system and its interpretation and the strategies to explore and test the configurations leading towards a solution for the problem situation.

The interaction between the mental activity and the physical activity is developed through seven stages according to Norman (Norman, 1986). The perception, interpretation and intention of physical activity is approached in relation to the problem-solving goal, action specification and execution at physical activity level. These different stages are monitored by metacognitive judgements and processes to adapt the learning strategy to solve the task at an object level.

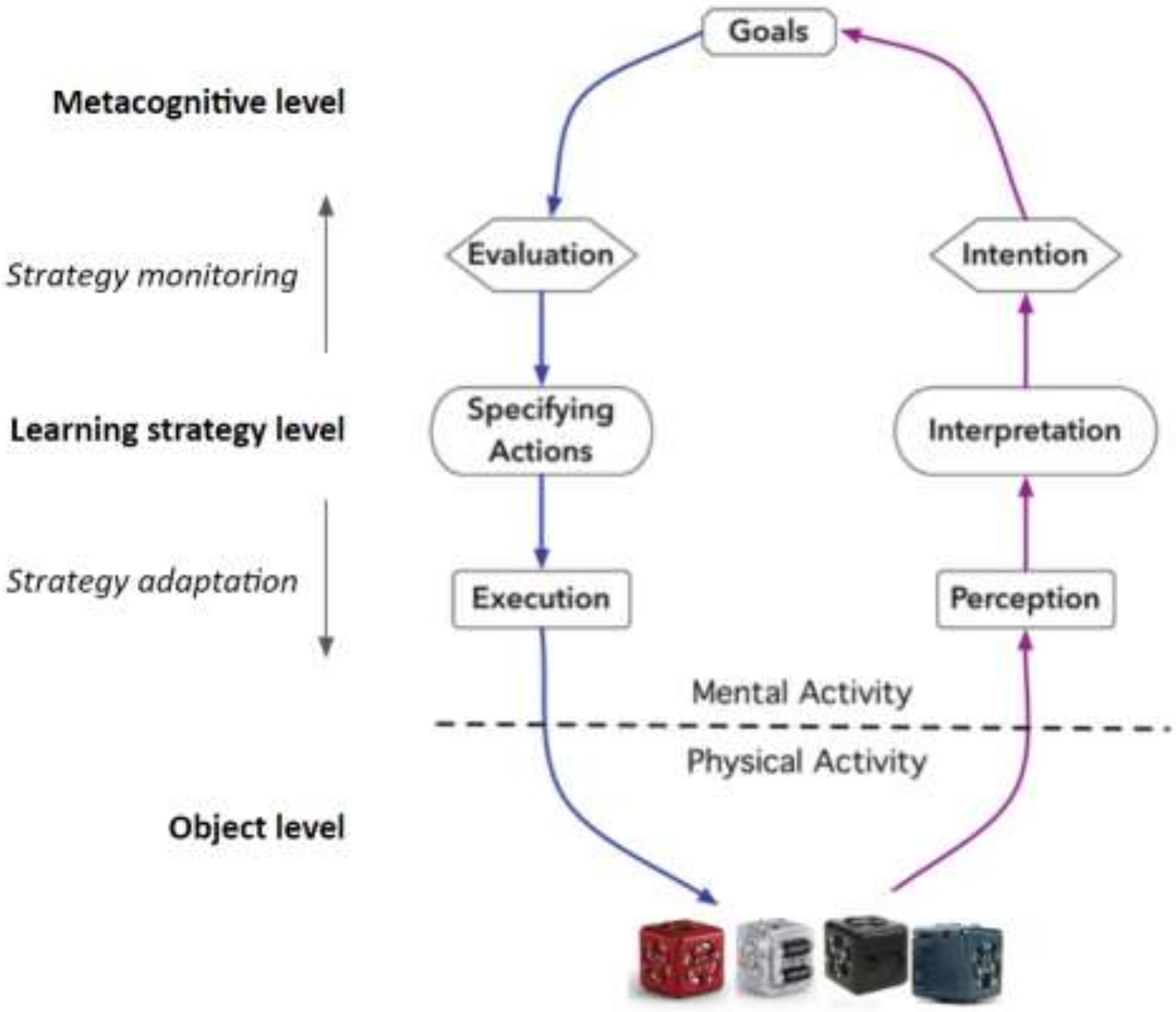

Figure 2 – Interactionist process for CPS based on Norman ( 1986)

## Creative Problem Solving with unfamiliar technologies

In CPS tasks with technologies that users are not familiar with, there is not the possibility of planning and executing an algorithmic solution; the gulf of execution of the proposed CPS task, and the lack of familiarity with the technologies proposed to participants requires exploring them to discover their potential affordances and refine both participants' knowledge as well as the state of a technological system. Through this whole process, a metacognitive regulation enables monitoring of the learner strategies and adaptation of these strategies towards the goal.

Within the process of interpreting the goals but also the technological system, the knowledge of the user is of key importance both in perceiving, interpreting, and evaluating the technological system. Being aware of the lack of prior knowledge related to unfamiliar technologies requires that participants interact with objects they want to use or create, in an emergent way, by acquiring and applying the technological knowledge required to solve the task. The manipulative exploration of the objects changes the problem space and contributes to constructing the technological knowledge necessary to move towards the solution space.

The actualization process of the technological system and the participants' knowledge is developed through the CPS process interactions and the metacognitive monitoring and regulation of these interactions. While users' actions change the state of the technological system, users' knowledge in relation to their interpretation of the technological system's feedback contribute towards actualizing users' knowledge.

In this context, technological knowledge can emerge from the interaction between participants and technology, impacted by prior knowledge and the emergent knowledge conceived through the exploration of the technology during the task. During this exploration of technology, physical affordances and functional affordances are also perceived by the user. "Perceived affordances and feedforward essentially provide different information about the action that users must perform to achieve their goals. While perceived affordances reveal the physical affordance, which tells users that there is a physical action available and how to perform it, feedforward reveals the functional affordance, which tells users what will happen when they perform that action" (Vermeulen and Luyten, n.d.). These interactions need to be monitored through metacognitive processes to ensure the way prior knowledge is actualized in relation to the interactions with the system and the value of this knowledge in relation to the goal of the problem-solving activity.

Participants can perceive a robot having "eyes" when seeing the distance sensors if they do not have prior knowledge of sensors (their appearance, their behaviour). The "eyes" can be considered as a perceived affordance for Vermeulen et al. (Vermeulen and Luyten, n.d.). When interacting with a robot, the robot can react when the participants' hands are in front of its "eyes". The participants can then construct a certain knowledge associating the gesture performed in front of the robot's "eyes" to a certain behaviour on the robot (functional affordance). Once participants have developed this technological knowledge, they can reuse it to solve the task. By acquiring technological knowledge through the task, the subject reduces the problem space and moves further into the solution space.

Technological knowledge can be unitary (e.g., a sensor component) or associative (multiple components assembled), when the technological objects are combined and generate new features or behaviours. We then distinguish the unitary-level technological knowledge and the system-level technological knowledge. In modular robotic technologies, technological knowledge should be constructed at a unitary level (each modular robotic component), but also considered as a system composed by different units having a specific system behaviour.

## Affordance perception, reasoning and metareasoning

In CPS with technology, users interact with a system to achieve a goal. The perception of the system is influenced by the way learners perceive the system's affordances. Affordances as a term was initially proposed by Gibson (Gibson, 1977) in order to describe object characteristics that are perceived both spatially but also due to their form by both humans and animals. Gibson's definition encompasses object characteristics that may or may not necessarily be perceived by animals or humans. The aspect of perception was also examined by Norman (2002), who made a distinction between real affordances, which are intrinsic to objects, and perceived affordances, which are related to what humans or animals perceive based on their interaction with objects. Like in other fields, affordances have an impact on the field of education. Kirschner et al. (2004) in their analysis, make a distinction between technological, social and educational affordances. While educational affordances encompass attributes of objects and content and their use in learning contexts, social affordances examine potential social and cultural interactions around learning contexts. Technological affordances refer to technological attributes and characteristics of platforms and tools.

The perception and interpretation of the affordances at object level (Figure 2) should be analysed at the interpretation level based on prior knowledge of the learner. Affordances exist as the interaction between the perception of stimuli and the prior experience of learners. Once affordances are interpreted and evaluated in relation to the CPS goal, the metacognitive activity adapts learning strategies regarding the interaction with the system towards a goal.

## Problem-solving as metacognitive and domain-specific activity

From an interactionist approach, this chapter models participants' knowledge and interactions but also the representation of the state of the robotic components across different tasks. The metacognitive level is examined in relation to the interpretation of the interaction with specific artifacts (robotic components) and the evaluation of the information in relation to the goal of the CPS activity and the adaptation of the learning strategies. In this section we firstly introduce the limits of the initial attempts to create a generic and universal approach to problem solving, to introduce the importance of domain-specific models focusing on specific tasks.

Problem solving has been approached through a variety of fields (Amado et al., 2018; Cai, 2003; Morfoniou et al., 2020). As a result, problem solving can be examined through the prism of game design (Kalmpourtzis, 2019), computational thinking (Romero et al., 2017, p. 21), design (Design Council, 2020), mathematics (Crespo and Sinclair, 2008) and business development (Ries, 2011). In order to examine the process of problem solving, various different processes have been proposed, including the ones by Polya (Polya, 1945), Dewey (Dewey, 1933), Torrance (Torrance, 1988). While these approaches to analyzing problem solving focus on the different phases that problem solvers can apply in order to arrive to a solution, the approach of Wallas (Wallas, 1926) focuses on the creative process that problem solvers pass through in order to generate solutions. On the other hand, creating problems, or problem posing, has been a topic of increasing research interest, due to its challenging and, on some occasions, demotivating nature towards students (Chen et al., 2011). In either case though, different sets of problem solving and problem posing skills have been identified by previous studies (Rosli et al., 2013).

The work of Newell, Shaw and Simon (Newell et al., 1959) aimed at developing a generic and universal approach to problem solving by considering humans as an information processing system that can operate according to modellable and representable rules. The General Problem Solver (GPS) presented by Newel in "Human Problem Solving" (Newell and Simon, 1972) has been a major contribution to cognitive psychology, more than in the field of artificial intelligence. In the criticism expressed by Ohlsson (Ohlsson, 2012) on the dead ends of the generic GPS approach, we identify the difficulty of understanding how subjects generate an initial problem space from a given situation, but also how complex situations are represented. Another aspect is intrinsic to the nature of the problem: the more a task is general, the less there is an effective method to solving it. This was identified since the 17th century, when Leibniz or Hobbes studied to what extent thought is reduced to a calculation (Tricot, 2017), and corresponds to a mathematically formalized result (Wolpert and Macready, 1997): any algorithm that would solve all problems could only be totally ineffective on each one of them individually (ultimately no better than just doing everything at random). In the implementation of the GPS model, independence did not allow it to be applied to different tasks. Current research on problem solving therefore emphasizes the importance of the task and the specific knowledge of the field to study problem solving, which should be approached through the framework of a specific task, while also considering prior knowledge of the subject and the context in which the task is developed. Sweller's work on problem solving and cognitive load (Sweller, 1988) underlines the difficulty of learning domain-independent heuristic strategies and recommends the learning of problems in specific tasks.

Despite clear evidence towards domain-specificity of CPS, we should consider that human cognition is developed based on a set of transversal skills such the task-generic metacognitive skills of planning, monitoring, and evaluation (Schuster et al., 2020).

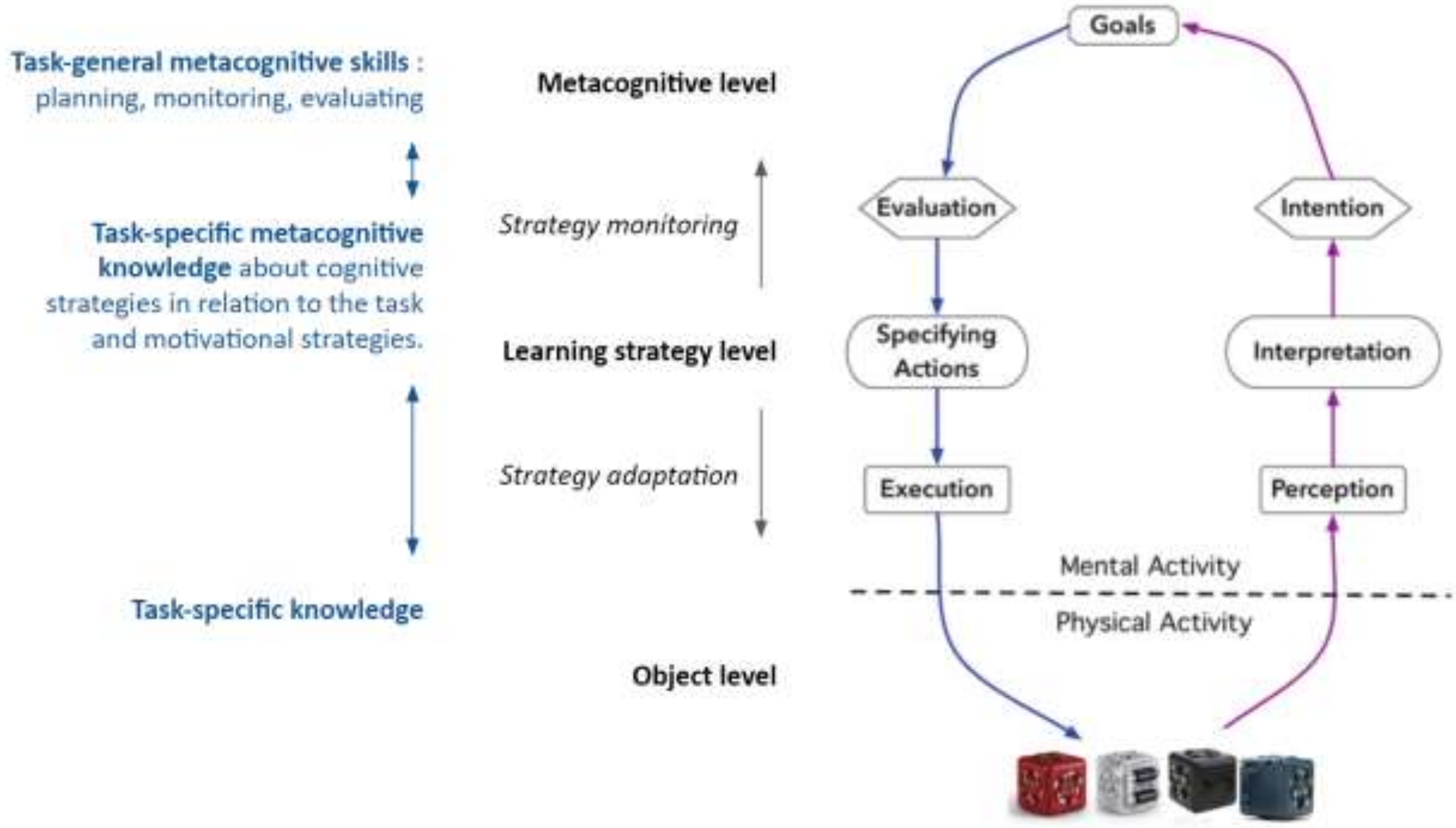

Figure 3 – Task-generic and task-specific metacognitive components in CPS

# Methodology

The aim of this study is to examine the approach problem-solving process with modular robotic technologies through an interactionist approach. This study presents a case study where an adult participant was presented with the CreaCube task. The sessions were run by a facilitator, who would arrange the cubes on a table and hide them using a blanket, while providing the audio instructions of the activity to the participants. The participant heard the same recorded instructions message and were presented with the same set of four different robotic modular cubes. The participant provided their informed consent for their participation in the task. The session was audio and video recorded.

## *Creative problem-solving modelling of the CreaCube task*

Considering the importance of considering domain-specific problem solving, this study models a creative problem-solving task with modular robotics. The CreaCube task presents participants with the goal of assembling and moving autonomously a set of modular robotic cubes from an initial point to a final point. To solve the task with a set of unfamiliar robotic cubes, the partici-pants first listen to a set of instructions (“You should build a vehicle made of four pieces moving autonomously from the red point to the black point”), and then observe and start manipulating the unfamiliar robotic cubes. Cube manipulation includes the gestures of touching and grasping the cubes to explore them individually. Cube manipulation will eventually lead participants to develop the technological knowledge associated with each unitary component: the movement potential of the drive cube having wheels, the energy potential of the battery cube, the distance sensing potential of the distance cube and the inverter signal potential of the inverter cube.

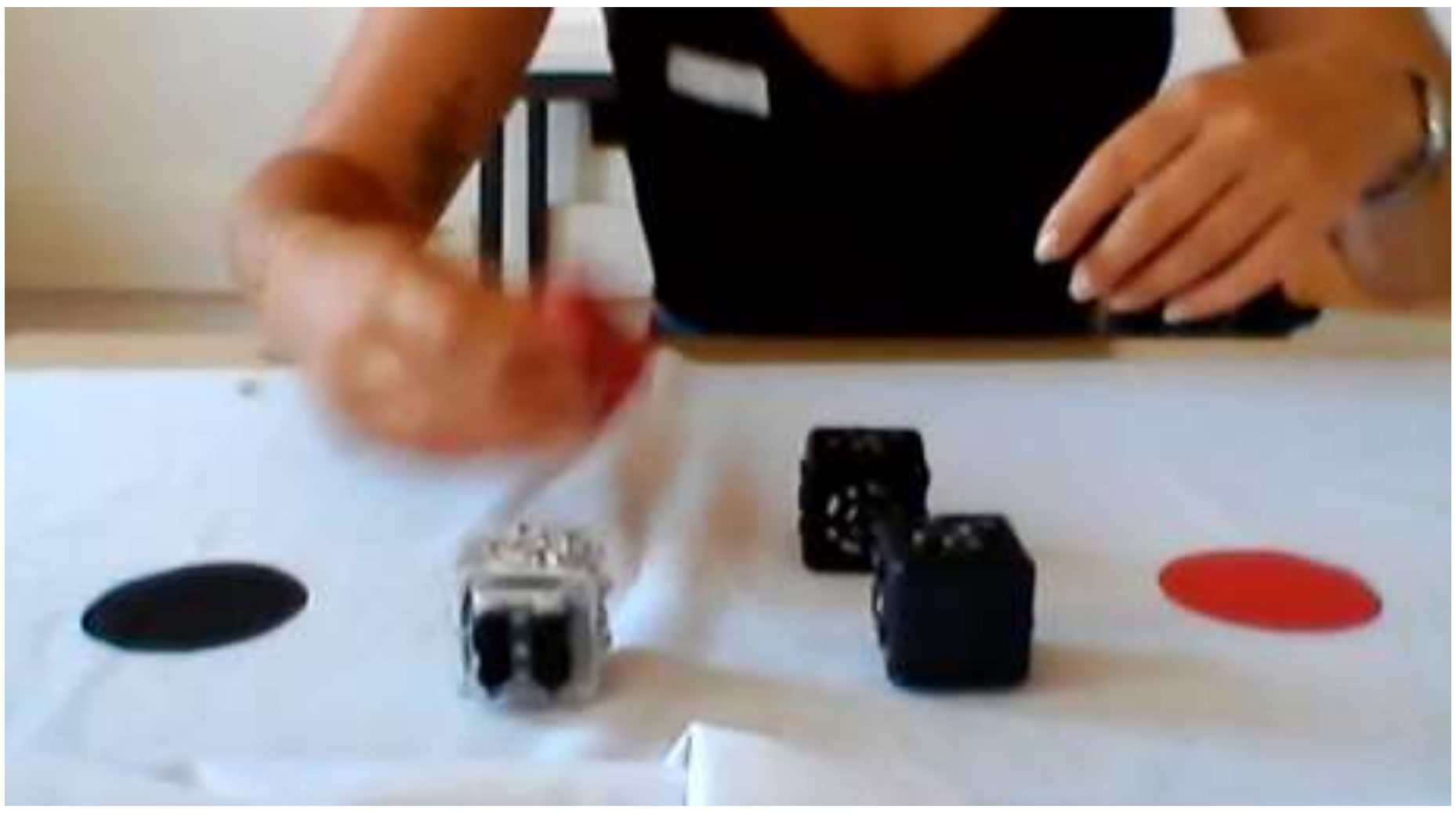

*Figure 4 – A participant engaged in the CreaCube task*

This potential technological knowledge is not always conceptualized through the manipulation of the cubes nor other stages of the task. While understanding the wheels and switching features are required to solve the task, some participants are expected to solve the task without a proper understanding of the sensor and the inverter cubes. Participants can assemble and disassemble the four cubes, hence creating different types of figures. Some of these combinations would potentially lead to solving the task (presented in green in Figure 5), some can solve the task in certain configurations (presented in orange in Figure 5), while others wouldn’t lead to a viable solution (presented in red in Figure 5).

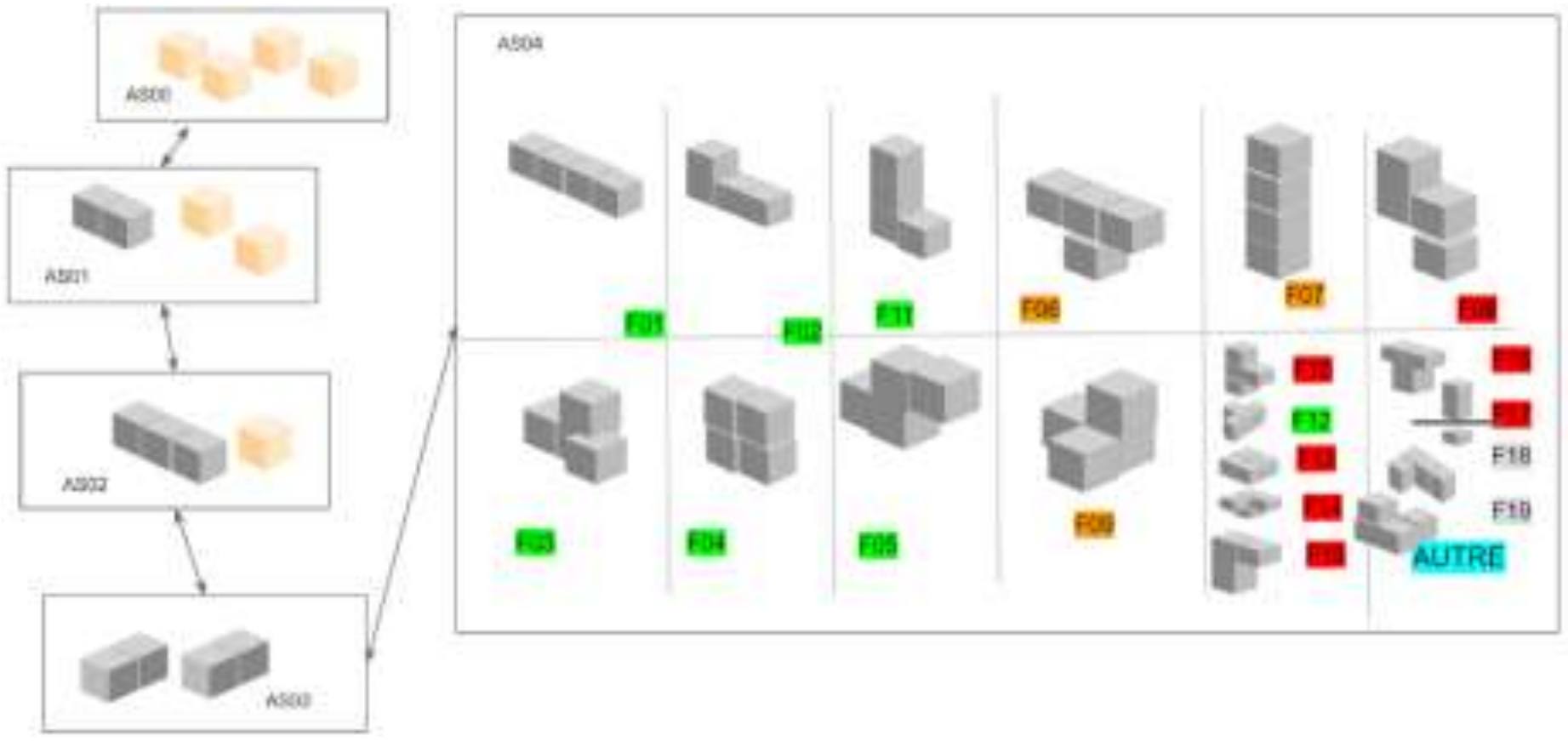

*Figure 5 - Different combinations leading or not to the solution of the task*

Each cube could be represented as having different states. When the state of each of the four cubes corresponds to the required solution state, the task is completed (the four assembled cubes will move from an initial point to a final point).

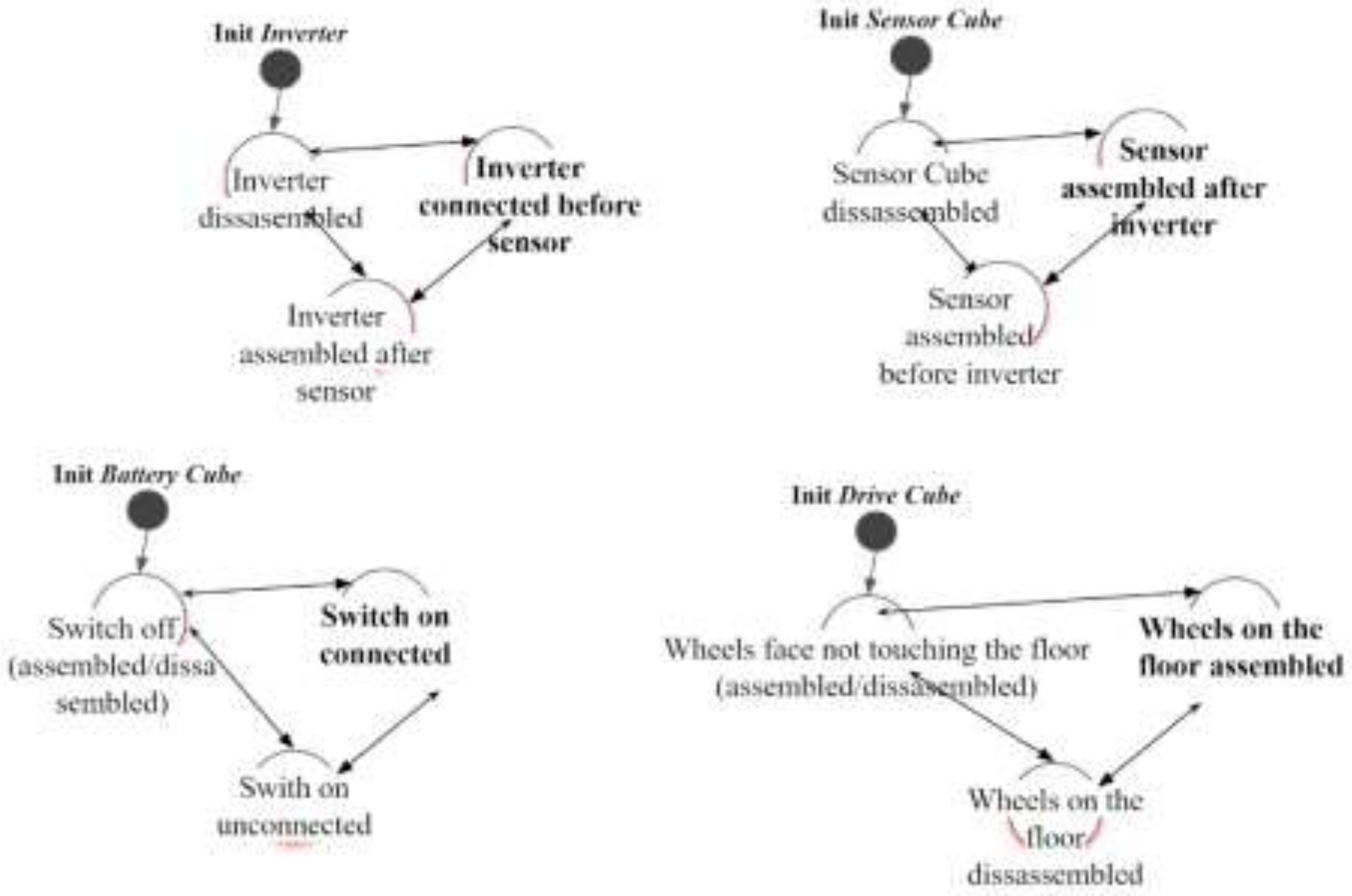

*Figure 6 - Modelling the user knowledge space across the CreaCube task*

## *Data collection and analysis*

This study presents a case study, where one participant joined a modular one-hour long CreaCube session. During this session, the participant initially listened to the audio instruc-tions. The session recordings were later analysed to identify how the participant approached a problem-solving task through an interactionist approach. For this reason, a qualitative research methodology was used. Initially one researcher watched through the entire session, coming up with initial labels, while the same researcher proposed axial coding based on a second pass of the video. A second researcher reviewed the results for triangulation purposes. The categories proposed were formed based on an iterative approach, where categories would be expanded or modified as the researcher would advance through the video recording.

## User knowledge space across the task

When engaged in the task for the first time, the participant did not have the technological knowledge needed to enable the required behaviour and location to succeed the task. The user should manipulate the cubes at the unitary level but also at the figure-system level to develop the technological knowledge required to complete the task. The exploratory interactions (see, touch, turn, assemble) the participant performed through the task facilitated the transition into the different states. All along the task the participant developed the mental model of the task based on the different inputs: hearing, seeing, touching, assembling. The first type of input are the instructions of the task the participant heard. The participant could listen to the instructions as many times she wanted, using a sound system (figure 7).

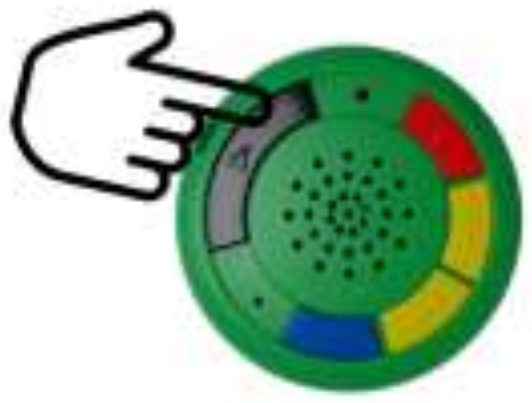

*Figure 7 - Sound system used to hear the instructions*

Based on the different prior knowledge of the participant and her mental state at the beginning of the study, the mental model developed when hearing the task instructions could be different. For some users, “vehicle” evokes a car or a train, for other persons, “vehicle” is defined as a set of features (a system with energy capable of movement).

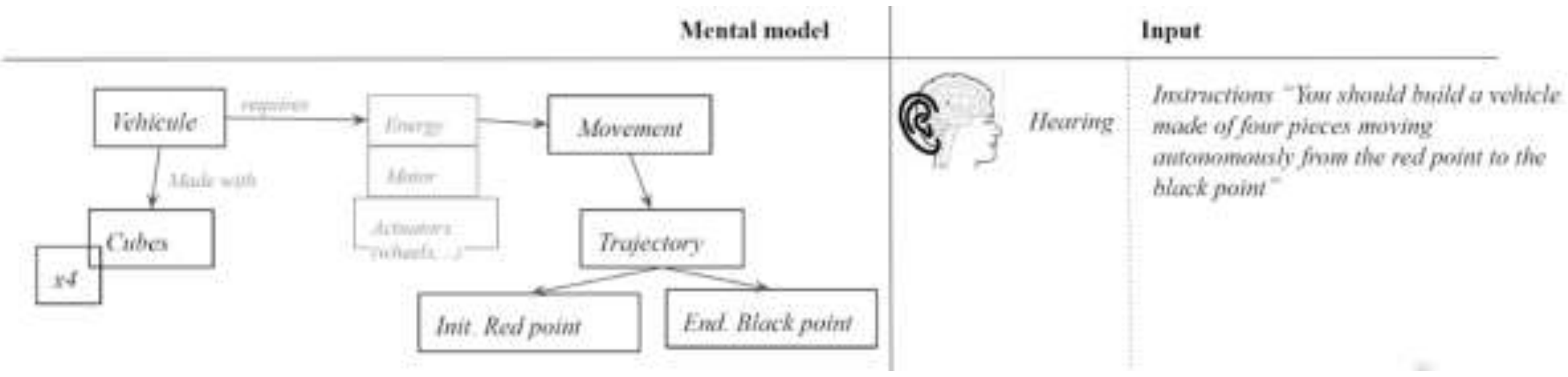

Figure 8 - Potential mental models while hearing the instructions

After the instructions were heard, the blanket covering the cubes was removed and the participant could see the materials: the four cubes and the two points. We expected the user to associate (in green) the visual input of the four cubes with the information of four cubes listening during the instructions, but also the red and black point in relation to the instructions. Moreover, the participant developed new knowledge (in blue): cubes have four different colours and metallic dots. Developing appropriate and relevant new knowledge contributes to the problem space reduction and going forward in the solution space.

We expect these affordances to lead to the interest in exploring the cubes further through the participant's orientation toward the means-goal of the task to make them move. Exploration is the key to perceiving the affordances (Berger & Adolph, 2007) and allows for the emergence of relevant knowledge about the material, which leads to use and to combining the material in such a way as to achieve the task goal.

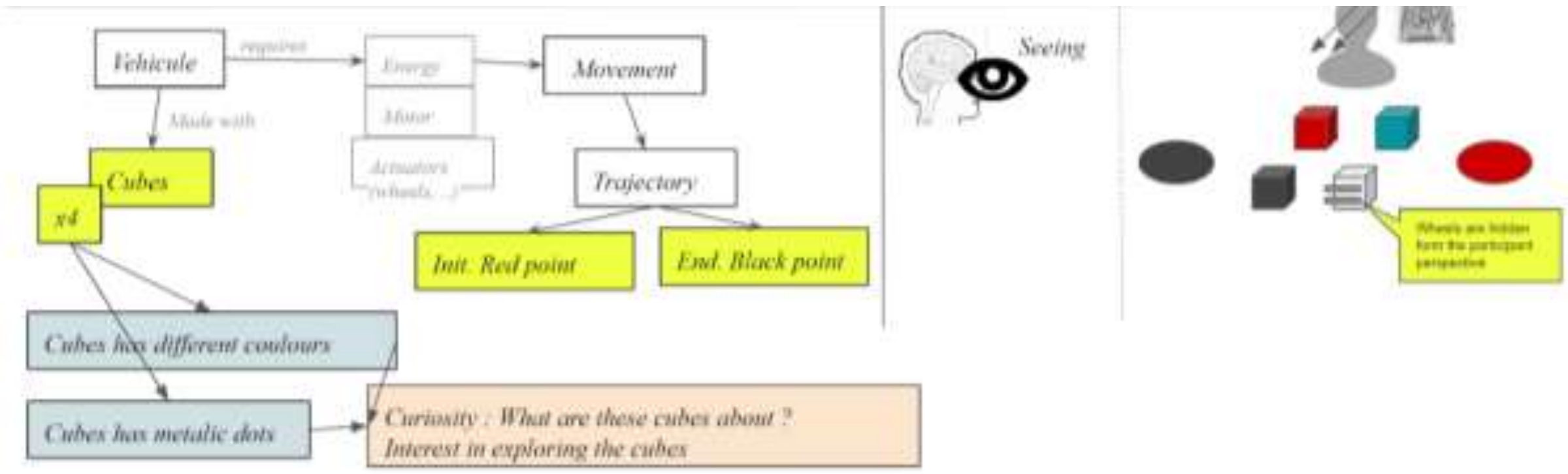

Figure 9 - Evolution of the mental model after seeing the material

After seeing the cubes, the participant grasped one or more cubes. While manipulating the cubes the participant could develop new technological knowledge in an emergent enactive way. Exploration of each cube facilitates the identification of the technological affordances.

We first describe the technological knowledge that emerges when interacting with the drive cube, which contains the wheel feature: the participant sees the cube with two black cylinders, which can lead to the concept of wheels. Touching the wheels can confirm they move as expected. By making the wheel roll on the floor the participant can observe that they move, but not autonomously or as a friction motor mechanism. If the reasoning is developed as expected, the participant should deduce that the wheels require a source of energy.

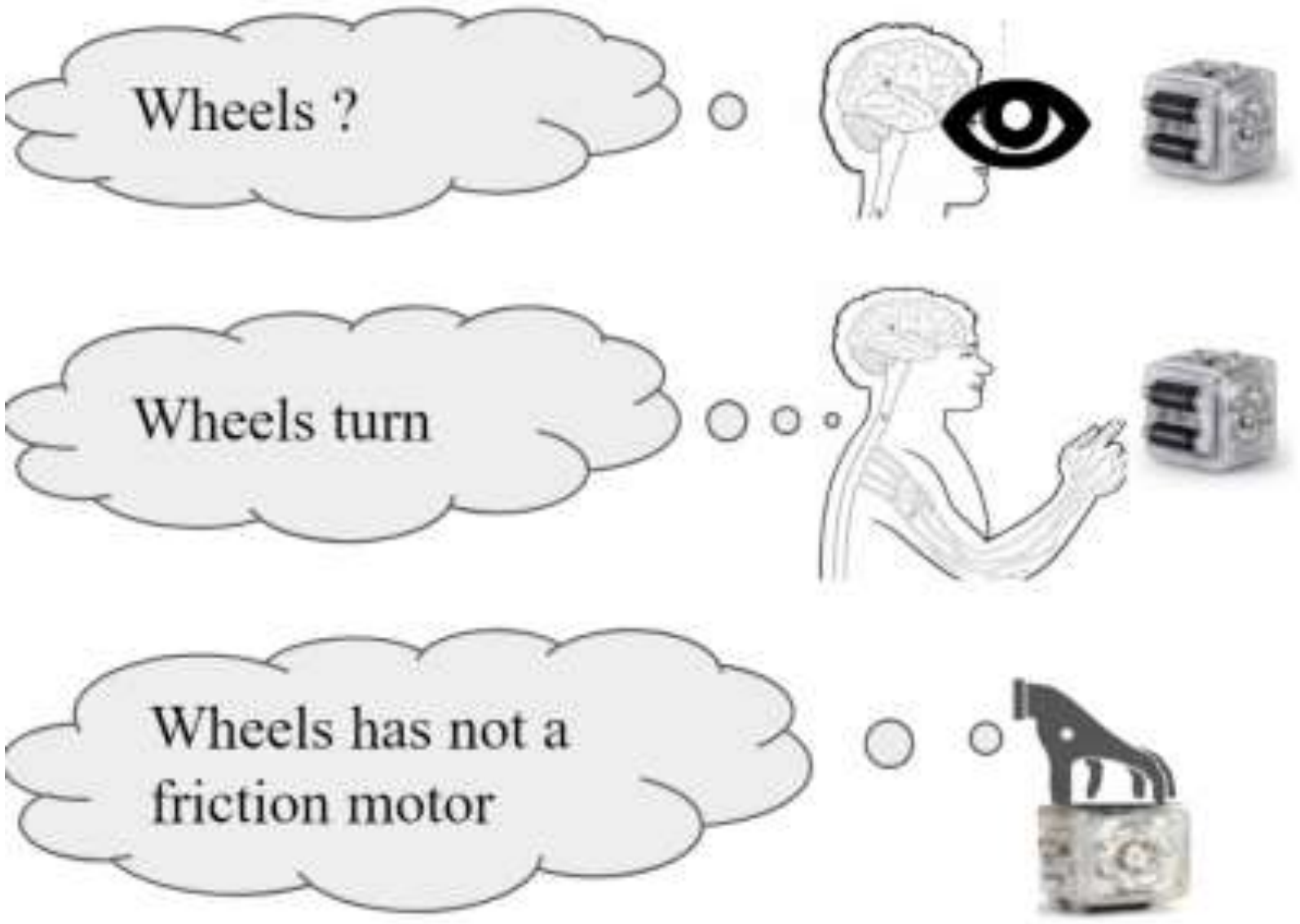

Figure 10 - Unitary exploration of the drive cube

When the subject perceives the wheels there is a construction process between the subjects' prior knowledge on wheels and a certain visual input corresponding to the concept of wheels by the subject. The perception is a construction process rather than a passive mechanism for processing incoming data which is determined by the subjects' sensory-motor capabilities, prior knowledge and culture and society in which the concepts are developed within the interactive loops of enaction we describe in this study.

The capacity to activity the appropriate prior knowledge during the task is important as highlighted by van Loon et al. ( 2013) which observed that children's "activation of inaccurate prior knowledge before study contributes to primary-school children's commission errors and overconfidence in these errors when learning new concepts" (p. 15). In the CreaCube task we have observed some of the participants to activate the prior knowledge of a "friction motor" with the wheels, which make them expect the work the same way as toy cars with friction motor which can advance after the toy is pushed forward (kinetic energy) and then released. Learners' which activate this prior knowledge test the material in this way and then they observe the system is not working as expected. While some learners develop a correct judgement of this prior knowledge and try other hypothesis just after trying once this hypothesis of a "friction motor", part of the learners require several tests to develop a more adapted judgement and discard this idea.

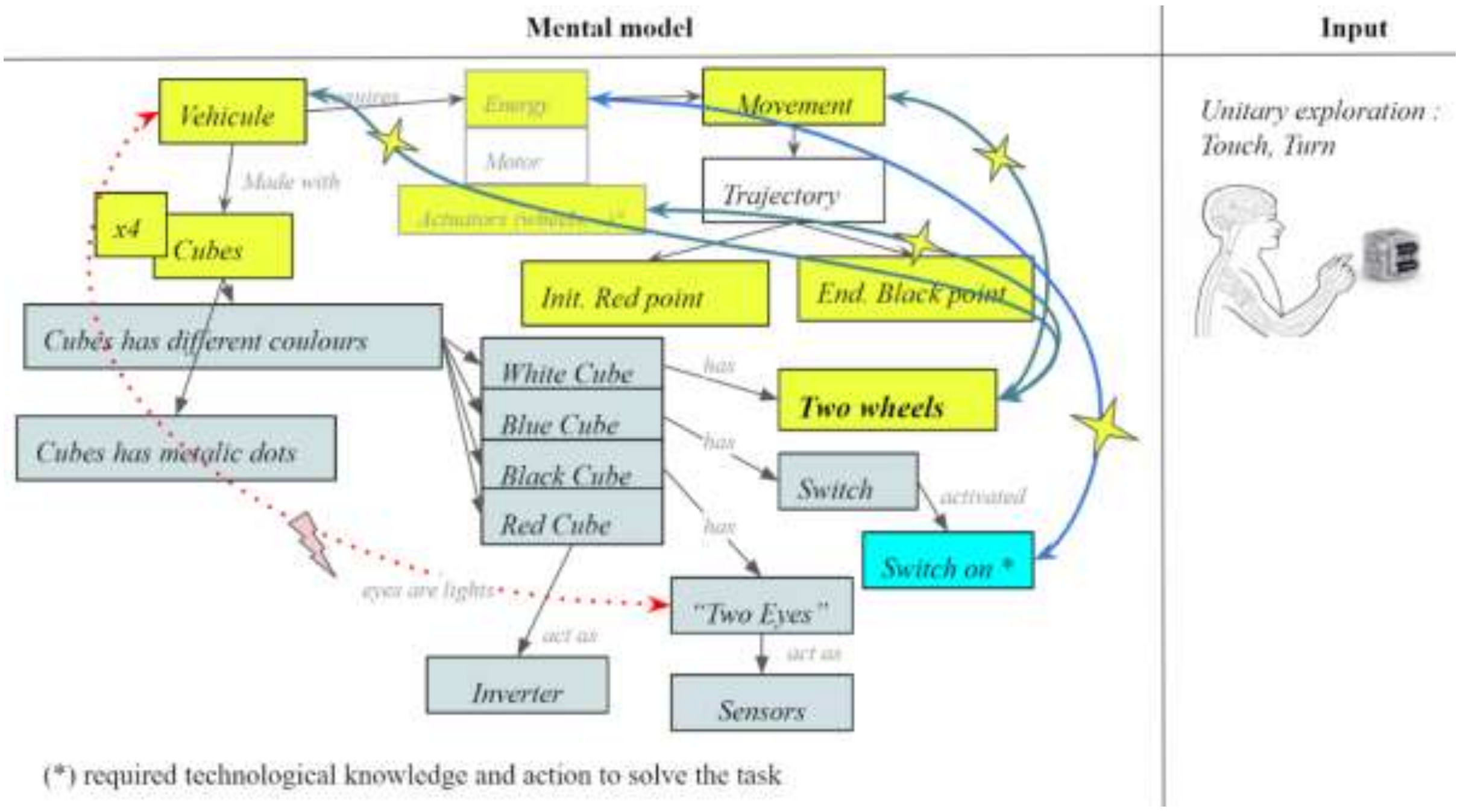

Figure 11 - Knowledge emergence during unitary exploration of the cubes

Knowledge emergence during the exploration is developed by the interaction between the learner and the material which requires to activate and then modify prior knowledge into a schema which could help to understand the actions towards the problem-solving goal. The CreaCube task permits to develop emergent concepts in a way the learner can solve the task. For example, a user which is not familiar with the concept of distance sensor will initially "perceive" the black cube as having "two eyes", which later the learner can figure out as a cube which reacts when the hand is in front of the "two eyes" and use this naïve or non-institutionalized knowledge as a schema to solve the problem without having the possibility to achieve the concept of "distance sensor" autonomously. For this reason, the debriefing after the CreaCube task is very important to help the learners' which have constructive naïve knowledge to institutionalize it in relation to the disciplinary knowledge of reference. Said another way, the learner with no prior knowledge of distance sensor can perceive "two eyes" reacting to his hand in front of the "two eyes" but will require a more knowledgeable other (MKO) (Abtahi et al., 2017; Vygotsky, 1980) to transform the germ cell of this concept developed during the activity into an institutionalized concept developed during the debriefing with the MKO.

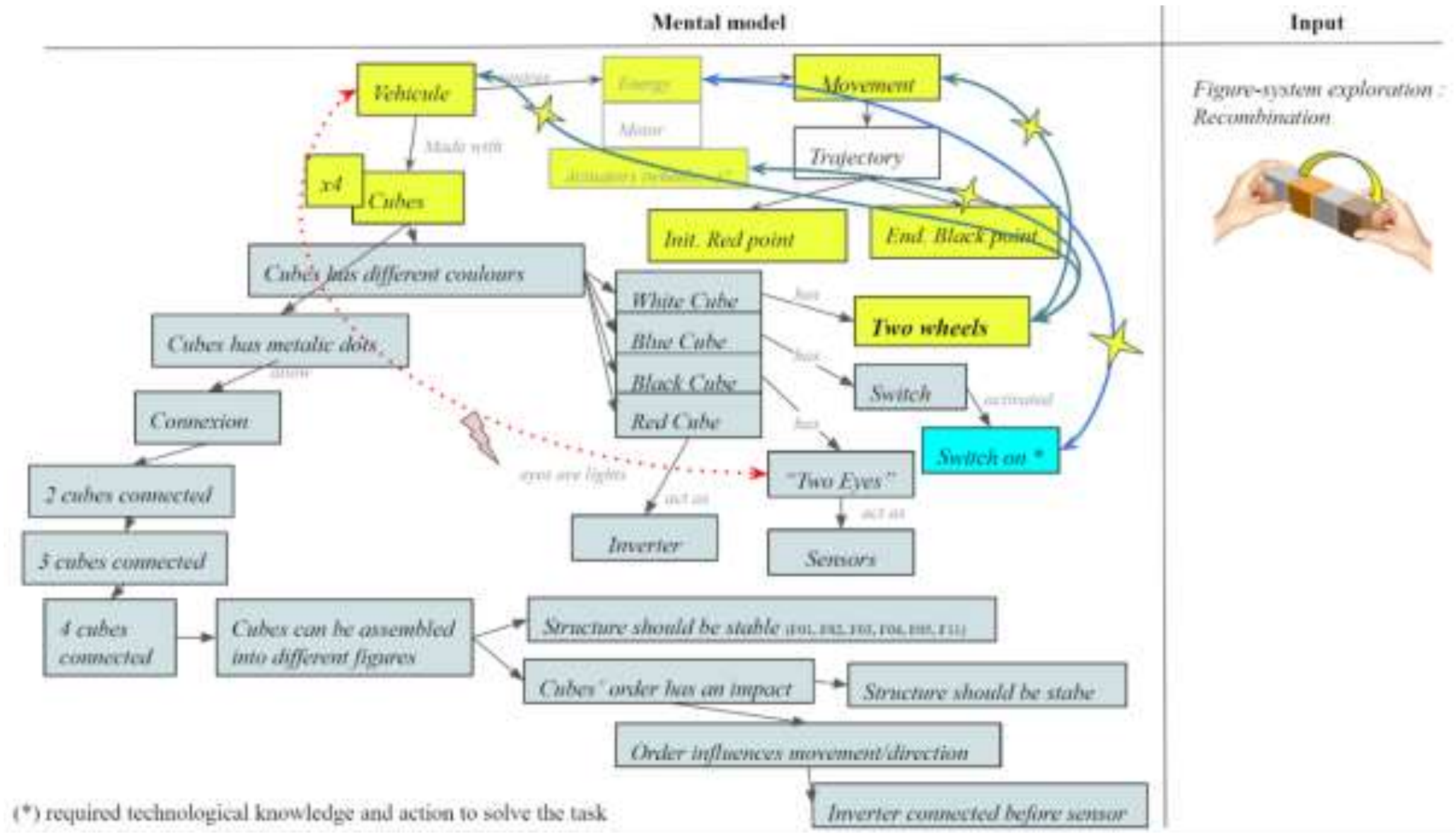

Figure 12 - Knowledge emergence during figure-system exploration and recombination of the cubes

## Metacognition judgements and regulation within creative problem

We have observed CPS within the CreaCube task to be a process which requires the learner to engage in a certain task and then regulate his creative CPS behaviour until reaching a satisfactory solution or dropping out the task. When we have candidate ideas that could help to solve a creative problem-solving task, we behave non creatively in a context in which we can engage in a creative behaviour. Creative behaviour requires not only a creative intention but also a creative behaviour perseverance (Leroy & Romero, 2021). When having the possibility to behave in a creative or non-creative way, when the subject identifies stable and predictable ideas or knowledge in relation to the task, these ideas or knowledge should be inhibited in the regulation of divergent and convergent thinking. In this process, the awareness of what is already known and what can be considered as new (metacognitive judgements on creative properties of originality, usefulness, and value) and a regulatory process allowing to inhibit certain ideas and strategies, and a deliberate process to generate new ideas (metacognitive monitor of creative divergent thinking strategies) and their selection according to the metacognitive judgements.

When considering creativity at the cognitive level, we should not only divergent thinking and convergent thinking but also the regulation of these processes. The regulation of divergent and convergent thinking could be considered as a type of meta control or regulatory processes of creative behaviour. The creative behaviour starts with a creative intention or volitional orienta-tion. During a CPS task the learner shows a certain creative intention, when showing a volitational orientation towards a creative behaviour during the task, and a creative behaviour perseverance when participants regulate their behaviour to maintain their creative intention across the task. Creative behaviour is considered to be part of system 2 reasoning (Kahneman, 2012), which is more slow, effortful and controlled than a conservative behaviour based on pre-existing knowledge and actions to perform a task (system 1, faster based on prior knowledge). Creative behaviour persistence requires a higher effort than conservative non creative behaviours. For achieving this perseverance there is a part of volitional orientation towards a creative process and outcome, but also a creative regulation including the metacognitive judgment and monitoring of divergent thinking and convergent thinking process, and therefore the regulation of explicit processing strategies of reflection (divergent thinking) or analysis (convergent thinking).
We consider the dual-process model of creativity of Augello et al. (2015) “which takes into account the different interaction mechanisms involving both S1 and S2 systems as well as both the generative and evaluative processes” [p. 7]. As shown in figure X, this model considers that some processes are implicit when generating ideas (exploratory process) but also when selecting them (tacit process), but other processes can be regulated explicitly at the divergent thinking stage (reflective process) and the convergent thinking one (analytic process).

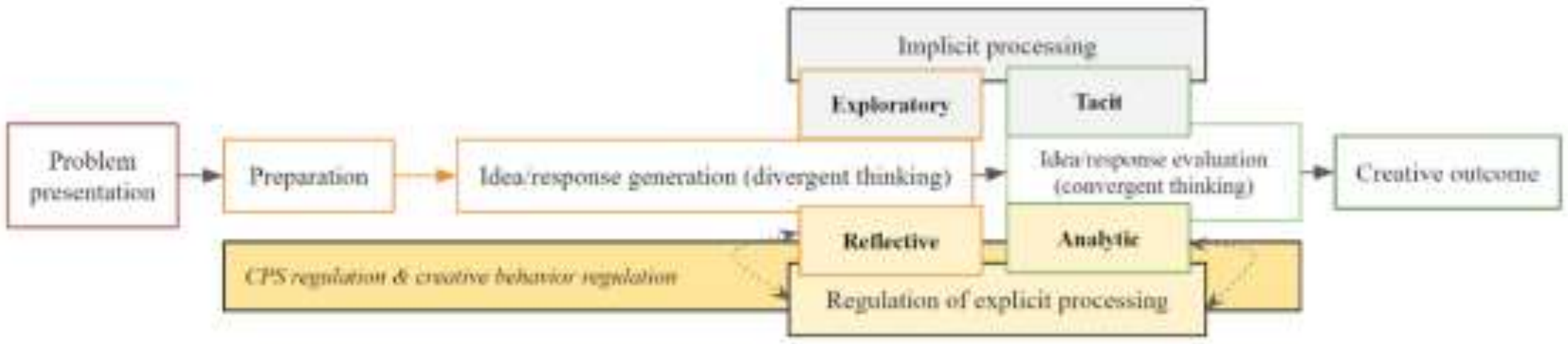

Figure 13 - Implicit processing and explicit processing in the context of CPS regulation.

When considering creative behaviour regulation in CPS we should then consider the level of the CPS task regulation and the first five phases (task level) and the level of the creative strategies (exploration/exploitation) and the phases (divergent/convergent) regulation (cognitive level regulation of creative behaviour).

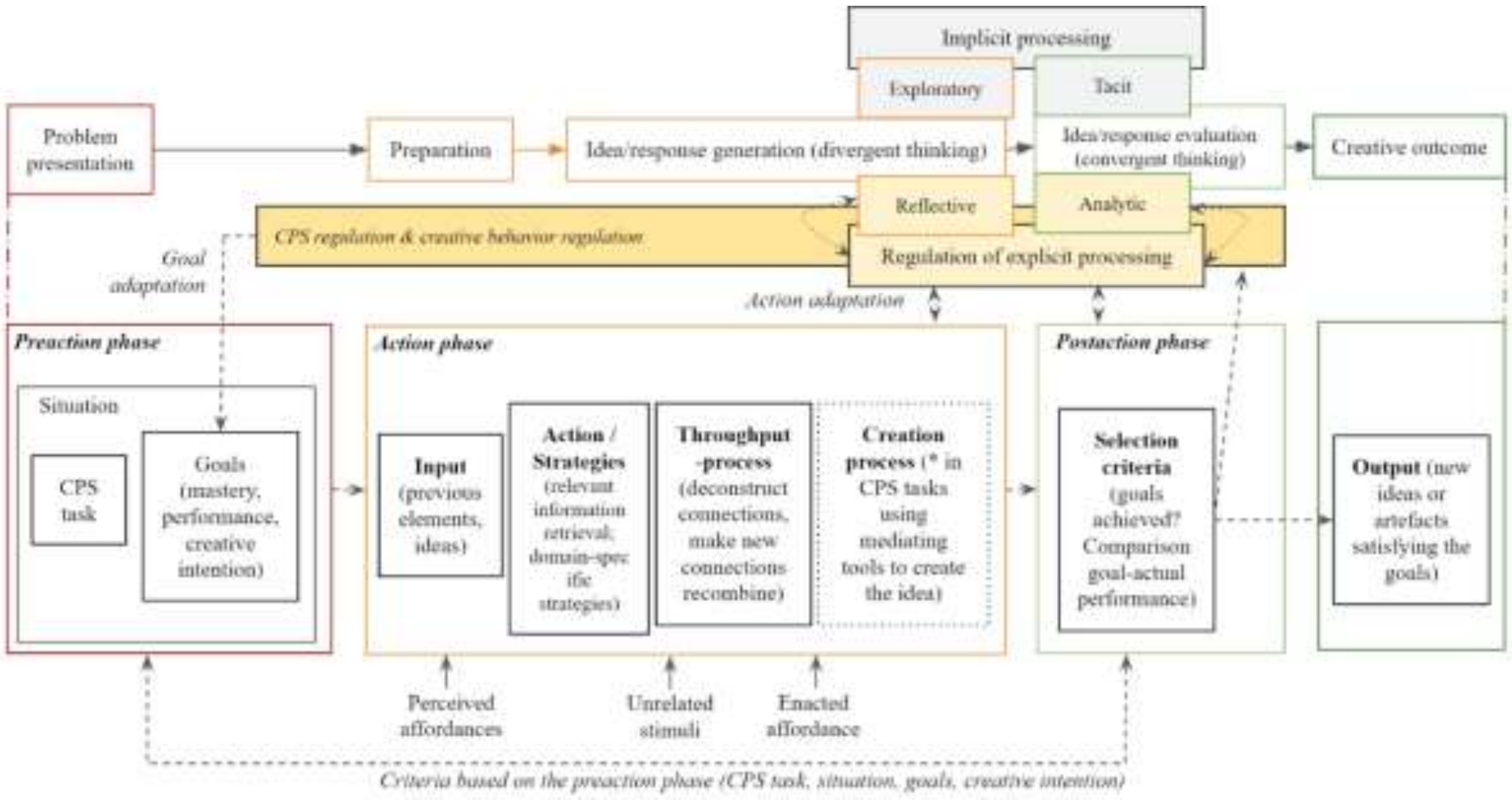

Figure 14 - Creative behaviour regulation in CPS (Romero, 2022)

Creative behaviour regulation in CPS (Romero, 2022) is multilevel (task, activity, cognition, metacognition and regulation). For the development of this model, we have considered the problem-solving regulation process of Perels et al. (2005), the phases of Amabile (1996) and integrating the dual-model or creative process of Augello et al (2016).

During the CPS, we consider volitional aspects and conflict of motives both from an activity system approach_(Sannino and Engeström, 2018) and cognitive approach (Di Domenico and Ryan, 2017) are crucial in the regulation process. When engaging in a CPS task, learners accept a “didactic contract” defined by the task instruction and context. However, the way learners engage in the task is not only defined by the task instructions, but also on the motivation of the learner in relation to the task goal and context. We consider motivation from a volitional perspective on which affects motivation and the capacity to regulate the creative process is also important.

## Discussion

This study contributes to a better understanding of solving with robotic technologies through an interactionist approach, in which participants’ prior knowledge, their initial representation of the task and their emergent construction of knowledge through the interaction with the material, is combined with the representation of the state of the robotic components across the task. We can observe the metacognitive level through the adaptation of the learning strategies through the CPS task. Through trial and error, the participant tried to propose different combinations that would provide a solution to the given problem. Throughout this process, the participant ventured a variety of cube combinations. By continuous exploration and testing and building upon her understanding during the affordance exploration phase, the solution space became narrower, based on the participant's previous experience. It was also observed that the participant, after identifying the affordance of some cubes, would capitalize on that understanding to explore potential affordances and their exploitation in their iterative process of creating a solution through assembling the four cubes into different configurations.

This study advances the modelling of metacognitive, and interactionists approaches of CPS with technological artifacts. In CPS, the learner should try to understand the problem, define problem solving strategies and apply them, while evaluating the outcome of her solutions through trial and error. The notion of affordances played an important role in the problem-solving process, since understanding the characteristics of each cube and what they afford impacted the final problem solution and the problem-solving process of the participant. As observed by An and Cao (An and Cao, 2014) metacognition is especially important in ill-structured problem-solving tasks, in which the learner has not all the required knowledge and procedures to solve the task beforehand. The CreaCube task allows us to observe the process of CPS and the different manipulative, cognitive and metacognitive levels which are required to solve an ill-defined problem in which learners should regulate their learning strategies to perceive the potential affordances, interpret them and actualize both the system they are manipulating, and the knowledge representations related to the task.

Through this study we have focused in a CPS task that could be used in formal and non-formal education context to evaluate divergent thinking (fluidity, flexibility and innovation) but also convergent thinking processes leading towards intermediate solutions and a satisfying creative outcome through the consideration of behavioural learning analytics in an interactive manipulative task with modular robotics (Mercier et al., 2021). The task can be used by the teachers to engage learners in a debriefing activity to consolidate not only the concept formation process during the activity (Romero et al., 2020; Sanchez and Plumettaz-Sieber, 2018) but also to engage the learners in the analysis of their metacognitive judgements during the task and the metacognitive activity which has permitted them to monitor and regulate their actions in the CPS task all the five phases (task level) and the level of the creative strategies (exploration/exploitation) and the phases (divergent/convergent) regulation (cognitive level regulation of creative behaviour). Moreover, the CPS model can permit to advance in the development of computational CPS models to simulate the learners' behavioural, cognitive, and metacognitive activity during the CPS permitting to develop studies which could make advance the research in the understanding of metacognition in CPS.

## Acknowledgements

This study was funded by the ANR Agence Nationale de la Recherche (ANR-18-CE38-0001).

Margarida Romero is full professor at Université Côte d´Azur in France and an associate professor at Université Laval in Canada. After starting her career at the Universitat Autonoma de Barcelona where she was awarded the best doctoral thesis in psychology, she continued her career in Canada and France, where she set up the Laboratoire d'Innovation et Numérique pour l'Education (LINE), a research unit in the learning sciences. She coordinates the #Scol_IA Working Group on AI in education and co-directs the MSc SmartEdTech program. Her research focuses on the study of computational thinking and creative problem solving.

George Kalmpourtzis is a researcher and consultant with an expertise in Human Computer Interaction applied to the Learning Sciences. He holds two BScs (one in education and one in computer engineering), a MSc in Advanced Information Systems and a PhD in Design Pedagogy and HCI. George helps organizations establish and improve learning strategies that improve employee and customer satisfaction. He has worked considerably on aspects like game-based learning, motivation and engagement and learning process creation. George also helps organizations create better product and service experiences through establishing design processes, user experience research and analytics.